\documentclass[12pt,preprint]{aastex}

\shorttitle{GK stars NLTE UV band}
\shortauthors{Short \& Campbell}

\begin{document}


\title{Modeling the near-UV band of GK stars, Paper III: Dependence on abundance pattern}


\author{C. Ian Short}
\affil{Department of Astronomy \& Physics and Institute for Computational Astrophysics, Saint Mary's University,
    Halifax, NS, Canada, B3H 3C3}
\email{ishort@ap.smu.ca}

\author{Eamonn A. Campbell}
\affil{Department of Astronomy \& Physics and Institute for Computational Astrophysics, Saint Mary's University,
    Halifax, NS, Canada, B3H 3C3}
\email{}





\begin{abstract}

We extend the grid of NLTE models presented in Paper II to explore variations in abundance 
pattern in two ways: 1) The adoption of the \citet{asplundgss09} (GASS10) abundances, 2) For stars of
metallicity, $[{{\rm M}\over{\rm H}}]$, of -0.5, the adoption of a non-solar enhancement of $\alpha$-elements by +0.3 dex.
Moreover, our grid of synthetic spectral energy distributions (SEDs) is interpolated to a finer numerical 
resolution in both $T_{\rm eff}$ ($\Delta T_{\rm eff} = 25$ K) and $\log g$ ($\Delta \log g = 0.25$).  We compare the values of $T_{\rm eff}$ 
and $\log g$ inferred from fitting LTE and Non-LTE SEDs to observed SEDs throughout the entire 
visible band, and in an {\it ad hoc} ``blue'' band.  We compare our spectrophotometrically derived $T_{\rm eff}$ values 
to a variety of $T_{\rm eff}$ calibrations, including more empirical ones, drawn from the literature.  For stars of 
solar metallicity, we find that the adoption of the GASS10 abundances lowers the inferred T$_{\rm eff}$ 
value by 25 - 50 K for late-type giants,  and NLTE models computed with the GASS10 abundances give 
$T_{\rm eff}$ results that are marginally in better agreement with other T$_{\rm eff}$ calibrations.  For 
stars of $[{{\rm M}\over{\rm H}}]$=-0.5 there is marginal evidence that adoption of $\alpha$-enhancement further lowers the derived 
$T_{\rm eff}$ value by 50 K.  Stellar parameters inferred from fitting NLTE models to SEDs are more 
dependent than LTE models on the wavelength region being fitted, and we find that the 
effect depends on how heavily line blanketed the fitting region is, whether the fitting region 
is to the blue of the Wien peak of the star's SED, or both.

\end{abstract}



\section{Introduction}

We have presented grids of LTE (\citet{shorth10}, Paper I) and Non-LTE (NLTE) (\citet{shorth12}, Paper II) atmospheric models and 
synthetic spectral energy distributions (SEDs) computed with PHOENIX \citep{hauschildtafba99} for GK dwarfs and giants (luminosity class V and III) of solar and $1/3$ solar 
metallicity ($[{{\rm M}\over{\rm H}}]=0.0$ and $-0.5$), and described a procedure for quantitatively comparing them to mean observed 
SEDs carefully selected from the large absolute spectrophotometric catalog of \citet{burn85} (henceforth, B85).  
Our strongest conclusion from Paper II is that the adoption of NLTE for many opacity sources 
shifts the spectrophotometrically determined $T_{\rm eff}$ scale for giants downward by
an amount, $\Delta T_{\rm eff}$, in the range of about 30 to 90 K all across the
mid-G to mid-K spectral class range, and across the $[{{\rm M}\over{\rm H}}]$ range from
0.0 to -0.5.  This shift brings our 
spectrophotometrically derived $T_{\rm eff}$ scale for the solar metallicity G giants into closer agreement
with the less model-dependent $T_{\rm eff}$ scale determined by the IRFM, although our $T_{\rm eff}$ 
values for these G giants are too large in any case.  
We also found tentative evidence on the basis of two spectral classes in the G range that this NLTE 
downward shift in the $T_{\rm eff}$ scale becomes smaller as luminosity class increases 
from III to V. 

\paragraph{}


In Papers I and II we divided the observed SED into two spectral bands, designated ``blue'' and ``red'', 
with a break-point at 4600 \AA, which is the wavelength around where the SED apparently changes character
qualitatively, from being
relatively smooth ($4600 < \lambda < 7500$, ``red'') to being more heavily affected by over-blanketing
by spectral lines ($3250 < \lambda < 4600$, ``blue'').  The quality of the fit of both LTE and NLTE model SEDs 
to the red band was significantly better than that to the blue band.  This is to be expected because 
the blue region is more heavily line blanketed, and is expected to be more sensitive to,
among other things, the distribution of the abundances of individual chemical elements.
This region is also relatively sensitive to inaccuracies and incompleteness in the input atomic
transition parameters (oscillator strengths, damping parameters, {\it etc.}), and to
the treatment of line formation and of the atmospheric models
({\it e.g.} 1D rather than 3D).  However, somewhat surprisingly, fits to the red and blue bands yielded
the same best fit $T_{\rm eff}$ values for many spectral classes. 

\paragraph{}

In the present investigation we take the step of 
perturbing the abundance distribution of our model grid, by perturbing both the fiducial solar abundance 
distribution adopted, and the choice of a solar or non-solar abundance distribution for the
models of $[{{\rm M}\over{\rm H}}]=-0.5$, and studying the effect upon the 
quality and model parameters of the best fits.  We believe this is timely and justified as there
have been important recent proposed revisions to the solar abundance distribution, as 
discussed in greater detail below (for example, compare \citet{grevs98} to \citet{asplundgss09}). 
A goal is to investigate the sensitivity of the spectrophotometric $T_{\rm eff}$ scale to 
the abundance pattern, and to compare it to the sensitivity to the thermodynamic treatment
(LTE {\it vs.} NLTE).  This will become increasingly important as instruments that can
produce calibrated spectrophotometry over broad wavelength ranges yield data for larger
samples of late-type stars ({\it eg.} VLT + XSHOOTER observations of red supergiants \citet{davies13}).  
Furthermore, accurate abundance measurements in late-type stars, particularly of refractory elements, has become
especially important when comparing planet-hosting stars to the general stellar population (see, for example,
\citet{adibekyands12}), and generally, the accuracy of abundance measurements depends on the accuracy
of the $T_{\rm eff}$ determination (see, for example, \citet{dobrovolskaska12}, who find from a NLTE analysis that 
${\rm d}[{{\rm Ba}\over{\rm Fe}}]/{\rm d}T_{\rm eff} = -0.03$ dex/80 K).


\section{Observed $f_{\lambda}(\lambda)$ distributions \label{sobsseds}}

In Paper I we described in detail a procedure for vetting and combining observed SEDs selected from the catalog of 
absolutely calibrated, and uniformly re-calibrated, spectrophotometry (SEDs) of 
B85, and we only briefly recapitulate here.  To state it more starkly than we did in Papers I or II, 
B85 is the largest catalog of absolute stellar spectrophotometry that we are 
aware of ($\sim 1500$ stars), and covers a broad enough $\lambda$ range (conservatively, $3300-7500 \AA$) to allow 
evaluation of
model SED fits to the heavily line-blanketed blue and near-UV spectral regions, given a fit to the
relatively unblanketed red region.  \citet{shorth09}
contains a more detailed description of the individual data sources
included in this compilation.  We emphasize again here, that this data set was uniformly {\it re}-calibrated,
by Burnashev (B85), so it should be less affected by random star-to-star errors than the original data.
Moreover, it contains large enough samples of stars of nominally 
identical spectral type (spectral class and luminosity class), as confirmed by cross-reference to other 
spectroscopic catalogs, and similar metallicity, as determined by cross-reference to the metallicity catalog of
\citet{cayrelsr01},
 that we can assess the variation, and thus the quality, of the SEDs and
reject those that are suspect.  The final results of our procedure are ``sample mean'' and $\pm 1 \sigma$ 
observed SEDs for each spectral type of each metallicity bin ($\Delta[{{\rm M}\over{\rm H}}]=\pm 0.1$) considered.  
Paper I contains details of how many SEDs were in each sample,
and plots showing the typical variation among the observed SEDs, for illustrative spectral
types.  These mean observed SEDs are the ``data product'' to which we compare our model SEDs here. 
 
\section{Model grid}

\subsection{Atmospheric structure and synthetic SED calculations}

\subsubsection{Basic grid \label{grid}}

  In Paper I we describe an LTE grid of spherical atmospheric models computed with PHOENIX 
for a star of mass $1 M_\odot$, and corresponding high resolution
synthetic spectra, that spans the spectral class 
range from about G0 to K4 ($6125 > T_{\rm eff} > 4000$ K, $\Delta T_{\rm eff} = 125$ K) and luminosity class range 
from V to III ($5.0 > \log g > 1.0$, $\Delta\log g = 0.5$),
and scaled solar metallicities, $[{{\rm M}\over{\rm H}}]$, of 0.0 and -0.5, 
with a mixing length parameter, $l$, defined by the B\"ohm-Vitense treatment of convection of one pressure scale height.  
In the current investigation, we are concerned with the sub-set of the grid corresponding to giants ($3.0 > \log g > 1.0$)
(the only metal-poor ($[{{\rm M}\over{\rm H}}]\approx -0.5$) GK stars in the B85 catalog are giants and supergiants, and 
the modeling of supergiants requires special additional considerations that are beyond the scope of this investigation.)
The choice of $[{{\rm M}\over{\rm H}}]$ values was guided by the availability of objects in the B85 catalog.
Paper II
describes the NLTE version of this grid, in which the lowest two ionization stages of 24 chemical elements up to Ni, 
and thousands of atomic transitions ($b-b$ and
$b-f$) are treated in self-consistent NLTE statistical equilibrium (SE), accounting simultaneously for NLTE radiative transfer
in those radiative atomic transitions that are accounted for in the SE rate equations.  

\subsubsection{Interpolated SED grid \label{Intergrid}}

In Paper I, we described a finer grid of SEDs, with $\Delta T_{\rm eff} = 62.5$ K, derived from this basic grid by linear interpolation
in the $\log f_\lambda(T_{\rm eff})$ relation.  Here we have derived an even finer grid, with $\Delta T_{\rm eff} = 25$ K and 
$\Delta\log g = 0.25$, 
by quadratic interpolation in the $\log f_\lambda(T_{\rm eff})$ and linear interpolation in the 
$f_\lambda(\log g)$ relations. 
The $\log f_\lambda(T_{\rm eff})$ interpolation was performed using two different methods to interpolate
values in the interval $[y_{i}, y_{i+1}]$: i) Simple quadratic interpolation by fitting a 
parabola to the three points 
{$y_{i-1}$, $y_{i}$, $y_{i+1}$}, and ii) Least-squares quadratic interpolation by fitting a parabola to the four points
{$y_{i-1}$, $y_{i}$, $y_{i+1}$, $y_{i+2}$}.  The two methods generally gave interpolated $\log f_{\lambda}$ values
that were within $1 \%$ of each other, and were both found to yield an interpolated 
$\log f_{\lambda}(T_{\rm eff})$ function that was smooth and well-behaved as determined by exhaustive visual inspection.  
Moreover, SEDs interpolated with both methods yielded the same best fit parameters to within the numerical
precision of our grid when the $\chi^{2}$ deviation from observed SEDs was minimized (see Section \ref{fitstats}). 
The results presented here are based on the least-squares quadratic interpolation.

\paragraph{}

The choice of $\Delta T_{\rm eff} = 25$ K is consistent with the recent ground-breaking investigation of \citet{huber12}, who found mean deviations 
of $22\pm 32$ K among values of $T_{\rm eff}$ derived with interferometric-asteroseismic, spectroscopic, and photometric techniques for stars 
of $4600 < T_{\rm eff} < 6200$ K, including red giants.  

\paragraph{Synthetic SEDs and microturbulence }

Synthetic SEDs suitable for comparison
to the observed SEDs are prepared by convolution with a Gaussian kernel with a FWHM value of 75 \AA, which was
found to be the value, to within $\pm 5$ \AA, that produced synthetic SEDs most closely matching the appearance of the observed SEDs.
The nominal sampling of the B85 catalog is 25 \AA, but the instrumental profile is apparently unknown.
In Paper I and II we presented a grid in which the microturbulent broadening parameter, $\xi_{\rm T}$, increases from 
2.0 to 4.0 km s$^{\rm -1}$ as $\log g$ decreases from 3.0 to 1.0.  To remove variation of a physically ill-motivated variable from our grid,
we have re-computed the atmospheric models and the synthetic spectra in this $\log g$ range with $\xi_{\rm T}=2.0$ 
km s$^{\rm -1}$.  Our tests show that for SEDs broadened for the observed spectral resolution element, $\Delta\lambda$, 
of 75 \AA, the difference between a SED computed with $\xi_{\rm T}=2.0$ and 4.0 km s$^{\rm -1}$ is completely negligible.   

\paragraph{}

\begin{figure}
\plotone{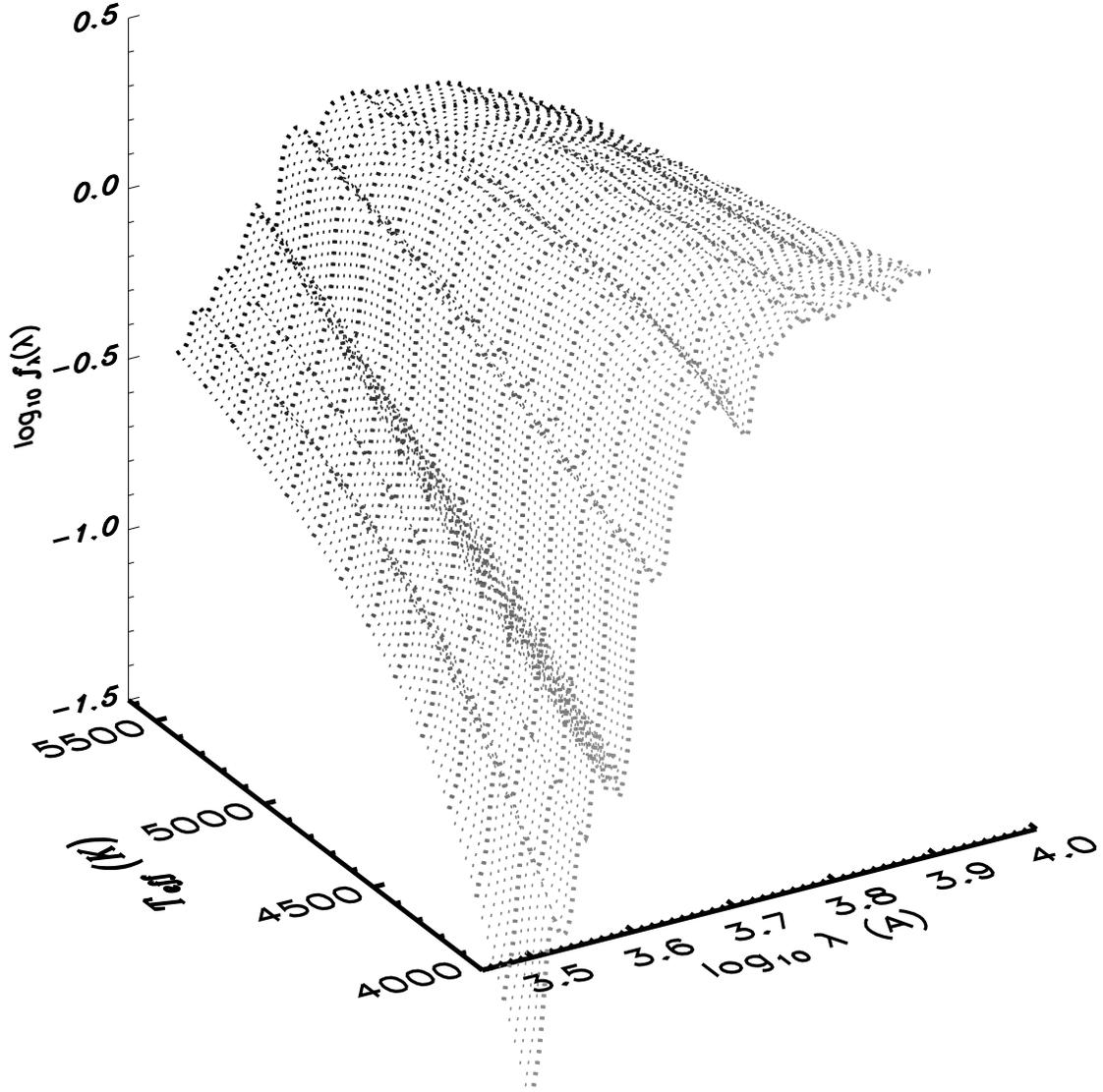}
\caption{The normalized, smoothed (see text) $\log f_\lambda(\log\lambda)$ surface as a function of $\log\lambda$ and model $T_{\rm eff}$ 
for the fine grid of solar abundance NLTE models with the abundances of GS98 and $\log g = 2.0$ interpolated
to $\Delta T_{\rm eff}=25$ K, for illustrative purposes.  The coarser
basic grid among which the finer grid was interpolated ($\Delta T_{\rm eff}=125$ K)
is also shown (thicker lines). 
  \label{refinterpolsurf}}
\end{figure}

Finally, both the observed and smoothed synthetic spectra are put onto the same flux scale by whole-area normalization 
following the procedure described in Paper II.
For illustrative purposes, Fig. \ref{refinterpolsurf} shows a plot of the interpolated $\log f_\lambda(\log\lambda, T_{\rm eff})$ surface,
broadened to match the spectral resolution of the data and normalized, as described above, 
for $\log g=2.0$ for NLTE models with the GS98 abundances.  The coarser grid from which this finer grid was interpolated
is also shown for reference.  We show the broadened $\log f_\lambda$ surface because the unbroadened surface is highly affected
by spectral line blanketing, and is visually confusing.  

\subsection{Abundances \label{sabund}}

\paragraph{Solar distributions }

The grid presented in Papers I and II was computed using the solar abundances of \citet{grevs98} (GS98, henceforth),
and this distribution was simply scaled for the $[{{\rm M}\over{\rm H}}]=-0.5$ models.  The GS98 
distribution is based on a compilation and review of abundances derived largely from 1D LTE semi-empirical solar photosphere 
and spectrum modeling, checked against meteoritic abundances for relevant elements.  These abundances, along with
other similar distributions published by the same authors over the years, have been widely used.  We have re-computed our 
$[{{\rm M}\over{\rm H}}]=0.0$ grid with the recent
abundances of \citet{asplundgss09} (see also \cite{grevass10}) (hereafter GASS10), which are based on recent 3D hydrodynamic solar photosphere and 
spectrum modeling, and which account for NLTE for some species, particularly C and O.  The GASS10 distribution has lower 
abundances, by $~0.1$ dex or 
more, than that of GS98 for the important elements C, N, O, Ne, Na, S, and K.  In particular, Na is among the 
most important $e^{\rm -}$ donors in the solar atmosphere.  Given the scale of the computational task of 
re-computing $\sim 500$ models and spectra, we have restricted ourselves to investigating the effects of
only these two solar distributions, and take them as being representative of the recent variation in the 
reputable measured solar abundance distributions.      

\paragraph{}

\begin{figure}
\plotone{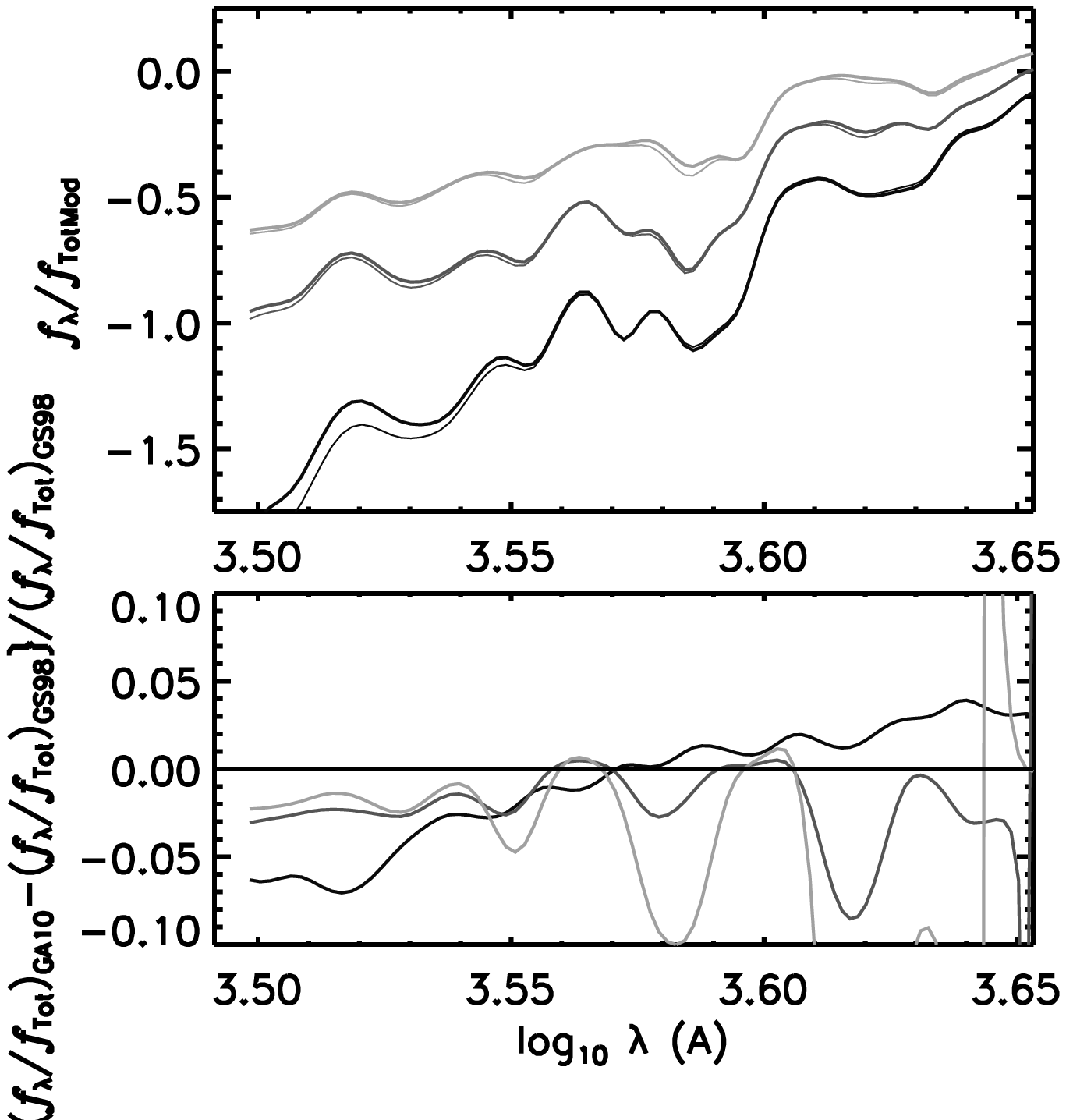}
\caption{Comparison of normalized, smoothed (see text) SEDs for the 3200 to 4500 \AA~ region
computed with different abundance distributions for models
of $[{{\rm M}\over{\rm H}}]=0.0$ and $\log g=2.0$.  Thin lines: GS98 models; thick lines:
GASS10 models; dark lines: $T_{\rm eff}=4000$ K; intermediate lines: $T_{\rm eff}=4625$ K,
lightest lines: $T_{\rm eff}=5250$ K.
Upper panel: Normalized SEDs (see text for description of normalization); lower 
panel: residuals (relative differences) with respect to the GS98 models.
Note that the residuals show large variation in regions where the flux derivative,
${\rm d}f_\lambda\over{\rm d}\lambda$, is large.
  \label{fcompspec1}}
\end{figure}

Fig. \ref{fcompspec1} shows a comparison, with residuals, of the area-normalized, smoothed NLTE SEDs 
computed with the GS98 and GASS10
abundances for solar abundance models of $\log g$ equal to 2.0 and $T_{\rm eff}$ equal to 4000, 4625, and 5250 K ({\it ie.}
spanning almost the entire $T_{\rm eff}$ range at a middle value of $\log g$).  (Note that the residuals
vary with large amplitude in $\lambda$ regions where the flux derivative ${\rm d}f_\lambda\over{\rm d}\lambda$
of the SED is large.)  The main effect of 
perturbing the abundance distribution is that the GASS10 abundances lead to a larger blue and near-UV band 
flux as compared that of the GS98 abundances, while the two distributions yield similar red band fluxes.
This effect becomes more pronounced as $T{\rm eff}$ decreases.
This is to be expected given that the GASS10 distribution generally has {\it lower} abundances than the GS98 distribution
for many relatively abundance ``light metals'' that contribute to the line blanketing, including Ti, V (both $\Delta[{{\rm M}\over{\rm H}}]=-0.07$),
Cr, and Ni (both $\Delta[{{\rm M}\over{\rm H}}]=-0.03$), 
thereby reducing the line extinction in the most heavily blanketed
spectral region (the blue and near-UV), thus allowing more flux to escape there.  This is essentially
a reduction in the classical line-blanketing/backwarming effect in the GASS10 models as compared to
the GS98 models.  Therefore, we expect that $T_{\rm eff}$ values inferred from fitting GASS10 models
will be {\it lower} than those of GS98 models to compensate for GASS10 models having a ``warmer'' SED at a 
given model $T_{\rm eff}$ value. 

\paragraph{Metal poor models}

 For our grid of $[{{\rm M}\over{\rm H}}]=-0.5$ models, we investigate the effects of four different 
abundance distributions: simply scaled solar distributions adopting the GS98 and GASS10 abundances, and
$\alpha$-enhanced distributions in which the scaled distributions of GS98 and GASS10 are altered
to a non-solar distribution by enhancing the abundance of eight $\alpha$-process elements (O, Ne, Mg,
Si, S, Ar, Ca, and Ti) by 0.3 dex.  We designate the latter two distributions GS98-$\alpha$ and
GASS10-$\alpha$.  Chemical abundance analyses of kinematically selected halo and 
thick disk stars in the solar neighborhood, including giants, have found that stars of 
$[{{\rm M}\over{\rm H}}]\le -0.5$, can have $\alpha$-element abundances enhanced by as much as 
$\sim 0.4$ dex (see \citet{adibekyands12}, \citet{adibekyanss12}, \citet{ruchti11}, \citet{fuhrmann11} and 
\citet{nissens10}, for examples based on large surveys and LTE analyses).  Furthermore, 
\citet{pdk93} performed a detailed LTE spectrum synthesis analysis of the high resolution spectrum
of the bright standard star Arcturus ($\alpha$ Boo), of $[{{\rm M}\over{\rm H}}]\sim -0.5$, and
found $\alpha$-element abundances enhanced by $\sim 0.3$ dex.  Recently, \citet{shi12} and \citet{takeda12} 
have carried out NLTE analyses of the abundance of the $\alpha$-elements Si and S, respectively, and confirmed similar
enrichment in moderately metal poor stars.  We have chosen an $\alpha$-enhancement
at the high end of the range that has been found for moderately metal-poor solar neighborhood stars ($+0.3$ dex), 
as exemplified by Arcturus
and applied it to all elements of even atomic number from O to Ti, to deliberately maximize
any effect on derived stellar parameters caused by $\alpha$-enhancement. 

\paragraph{}

\begin{figure}
\plotone{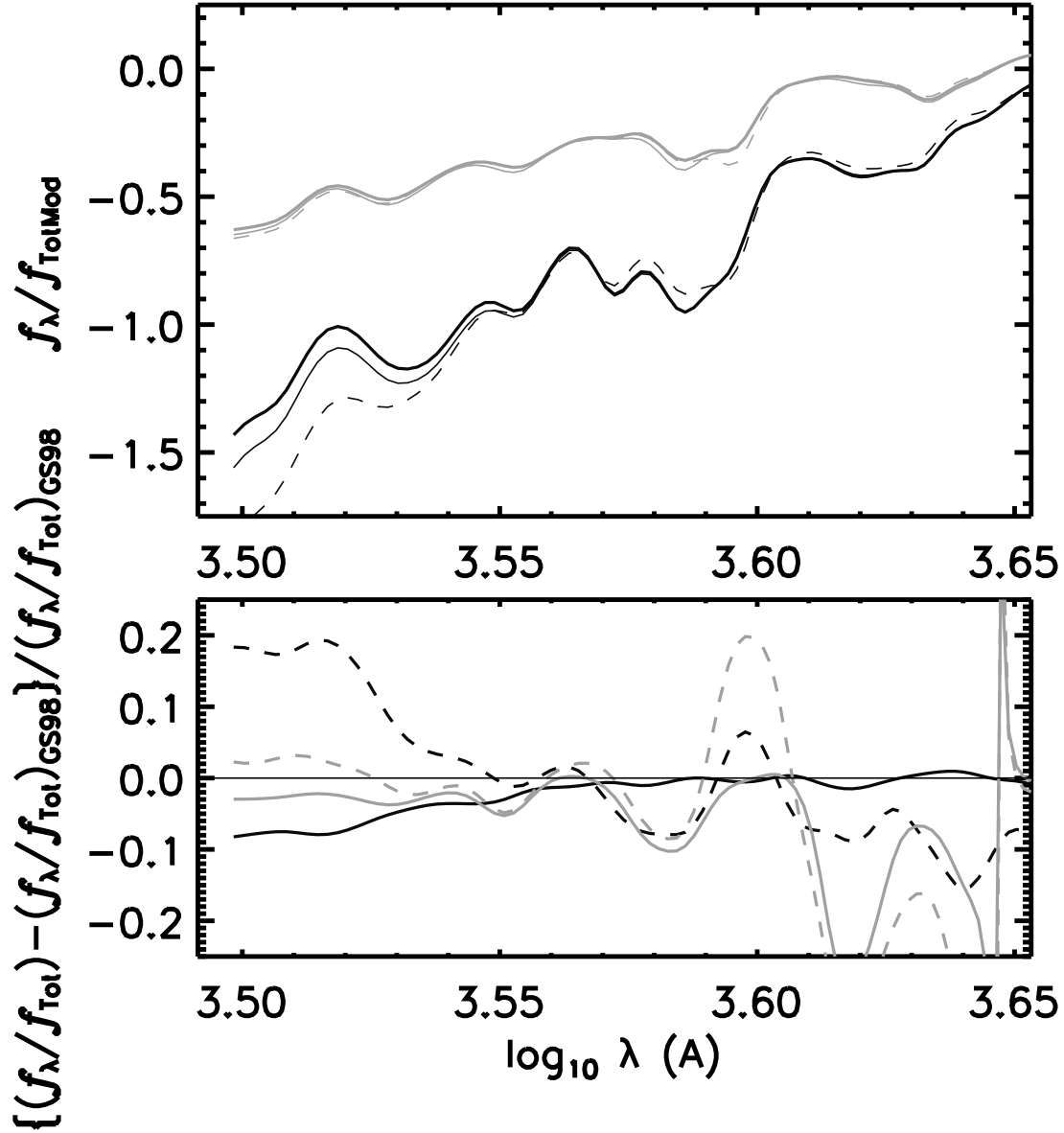}
\caption{Same as Fig. \ref{fcompspec1}, except for models of $[{{\rm M}\over{\rm H}}]=-0.5$,
at $T_{\rm eff}=4000$ K (dark lines) and 5000 K (light lines).
For the GS98 models, we also include SEDs computed with $\alpha$-enhancement (dashed lines).
In the lower panel all residuals are with respect to the GS98 models {\it without} $\alpha$-enhancement.
  \label{fcompspec2}}
\end{figure}

Fig. \ref{fcompspec2} shows a comparison, with residuals, of NLTE SEDs computed with the GS98 
and GASS10 abundances for metal-poor models at a $\log g$ value of 2.0, and $T_{\rm eff}$ values
of 4000 and 5000 K (similarly to Fig. \ref{fcompspec1}).  The blue/near-UV brightening caused
by adoption of the generally lower GASS10 abundances is again evident, despite the reduced significance 
of line extinction in moderately metal-poor models.  We expect the same reduction in the inferred
best fit $T_{\rm eff}$ value as for the solar abundance case, but with reduced magnitude. 
For the GS98 models, we also show the SED with and without 
$\alpha$-enhancement.  The effect of $\alpha$-enhancement is
to decrease the blue/near-UV band flux with respect to the red band flux, yielding 
``cooler'' SEDs for a given model $T_{\rm eff}$ value.  We expect $\alpha$-enhancement
to have the same effect as changing from the lower GASS10
abundances to the higher GS98 abundances in that it should lead to ``cooler'' SEDs for a 
given model $T_{\rm eff}$ and therefore, a larger best-fit $T_{\rm eff}$ value.

\section{Goodness of fit statistics \label{fitstats}}

  We minimize $\chi^{2}$ for the difference between the observed SED and the trial synthetic SEDs (the ``hypotheses''), relative to the observed 
SEDs ({\it ie.} ($f_{\lambda, {\rm Obs}}-f_{\lambda, {\rm Mod}})/f_{\lambda, {\rm Obs}}$) in each wavelength sampling element, $\Delta\lambda$.  This is equivalent 
to adopting unity ($f_{\lambda, {\rm Obs}}/f_{\lambda, {\rm Obs}}$) as the ``observed'' quantity, and ($f_{\lambda, {\rm Mod}}/f_{\lambda, {\rm Obs}}$) as the
trial quantity; {\it ie.}  

\begin{equation}
 \chi^2 \Delta\lambda^{\rm -1} = {1\over N}\sum_\lambda^N ((f_{\lambda, {\rm Obs}}-f_{\lambda, {\rm Mod}})/f_{\lambda, {\rm Obs}})^2 / (f_{\lambda, {\rm Mod}}/f_{\lambda, {\rm Obs}})
\end{equation}

or, equivalently,

\begin{equation} 
  \chi^2 \Delta\lambda^{\rm -1} = {1\over N}\sum_\lambda^N (1.0 - f_{\lambda, {\rm Mod}}/f_{\lambda, {\rm Obs}})) / (f_{\lambda, {\rm Mod}}/f_{\lambda, {\rm Obs}}) 
\end{equation}

where N is the number of wavelength elements minus 1, which we take to be the number of degrees of freedom.  The sampling interval, 
$\Delta\lambda$ is 15 \AA, which was chosen to slightly over-sample SEDs broadened to match a spectral resolution element of 75 \AA~ 
(see Section \ref{Intergrid}).
For each trial model we compute three values of $\chi^2 \Delta\lambda^{\rm -1}$,
one each for  our ``blue'' band (3250 - 4600 \AA), our ``red'' band (4600 - 7500 \AA), and the entire range, to study the quality of the fit to the 
over-blanketed blue region with respect to the red region.  For fits to the overall spectrum, the minimum value of 
$\chi^2 \Delta\lambda^{\rm -1}$ ranged from 0.002 to 0.005, with the exception of the coolest, most line blanketed sample, K3-4 III,
for which the $\chi^2 \Delta\lambda^{\rm -1}$ minimum was $\approx 0.01$.  These values are also characteristic of the fit to the
``blue'' band alone, indicating that the quality of the fit to the more heavily line-blanketed blue band dominates the quality of the
global fit.

\paragraph{}

Paper II contains a detailed comparison of the stellar parameters derived from fitting NLTE model SEDs as compared to LTE SEDs, and we do not 
repeat that discussion here, except to note that the main result was that NLTE $T_{\rm eff}$ values derived from fitting the overall visible band 
were found to be systematically lower than LTE values by one numerical resolution element, $\Delta T_{\rm eff}$, in the coarser SED grid of Paper I
(nominally 62.5 K).  With the 
finer numerical resolution of our current SED grid, we find that this $T_{\rm eff}$ difference for the GS98 models generally varies from 50 to 75 K
(in the new grid, this amounts to two to three numerical resolution elements, $\Delta T_{\rm eff}$). 
This is a systematic offset between the NLTE and LTE SED-derived $T_{\rm eff}$ scales that we are now numerically resolving, whereas in 
Paper II it was only marginally resolved.   

\paragraph{}

\begin{figure}
\plotone{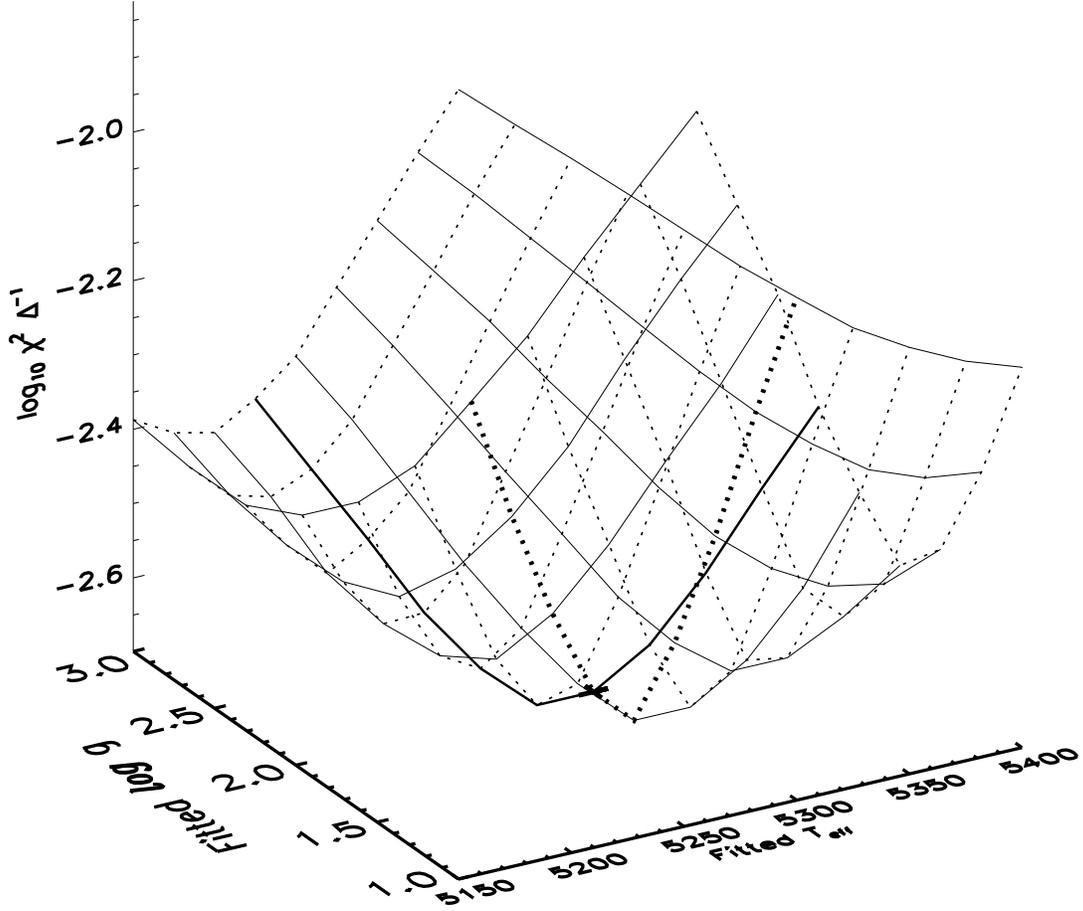}
\caption{The $\log\chi^{2} \Delta\lambda^{\rm -1}$ surface as a function of trial $T_{\rm eff}$ and 
$\log g$ values in the vicinity of the $\chi^{2}$ minimum for the G5 III/$[{{\rm M}\over{\rm H}}]=0.0$ 
sample and NLTE models of GS98 abundances.  Loci of constant $T_{\rm eff}$ and $\log g$ are shown
with solid and dotted lines, respectively, and the model $T_{\rm eff}$/$\log g$ pair of minimal
$\chi^{2}$ value is marked with a '+' symbol.  The loci of of constant $T_{\rm eff}$ and $\log g$
corresponding to the best fit $\log g$ and $T_{\rm eff}$ values, respectively, are plotted with  
thicker lines.  We plot $\log\chi^{2}$ to enhance the weaker variation of $\chi^{2}$ with
model $\log g$ as compared to its variation with $T_{\rm eff}$.  
  \label{refG5surf}}
\end{figure}

\begin{figure}
\plotone{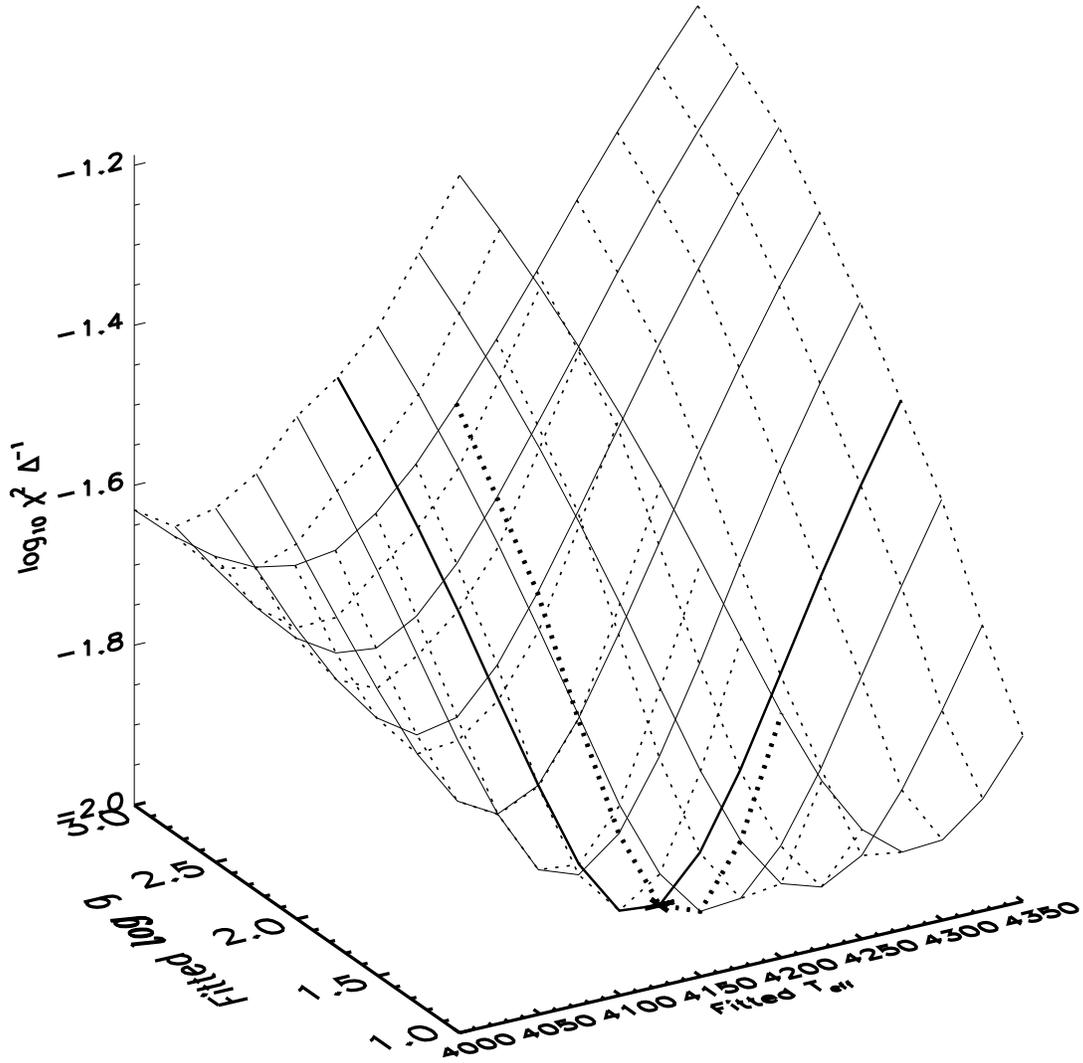}
\caption{The same as Fig. \ref{refG5surf} but for the K3-4 III/$[{{\rm M}\over{\rm H}}]=0.0$ sample.
  \label{refK3-4surf}}
\end{figure}

Figs. \ref{refG5surf} and \ref{refK3-4surf} show the logarithm of the $\chi^2 \Delta\lambda^{\rm -1}$
value as a $2D$ surface plotted {\it versus} model $T_{\rm eff}$ and $\log g$ in the vicinity of the $\chi^2$ 
minimum for the G5 III and K3-4 III samples of 
$[{{\rm M}\over{\rm H}}]=0.0$, respectively, for NLTE models with the GS98 abundances.  We have chosen a logarithmic 
scale to enhance the relatively weak variation of $\chi^2$ with $\log g$ with respect to its variation 
with $T_{\rm eff}$.  Visual inspection of such plots for all our spectral class samples assures us that
the $\chi^2$ minimum is global and well defined as a function of $T_{\rm eff}$.  The minimum is less
well-defined as a function of $\log g$, and in every case we have checked for local degeneracy in $T_{\rm eff}$/$\log g$
around the $\chi^2$ minimum (see the discussion of the K1 III/$[{{\rm M}\over{\rm H}}]=0.0$ sample in this
section).
  
\paragraph{}

\begin{figure}
\plotone{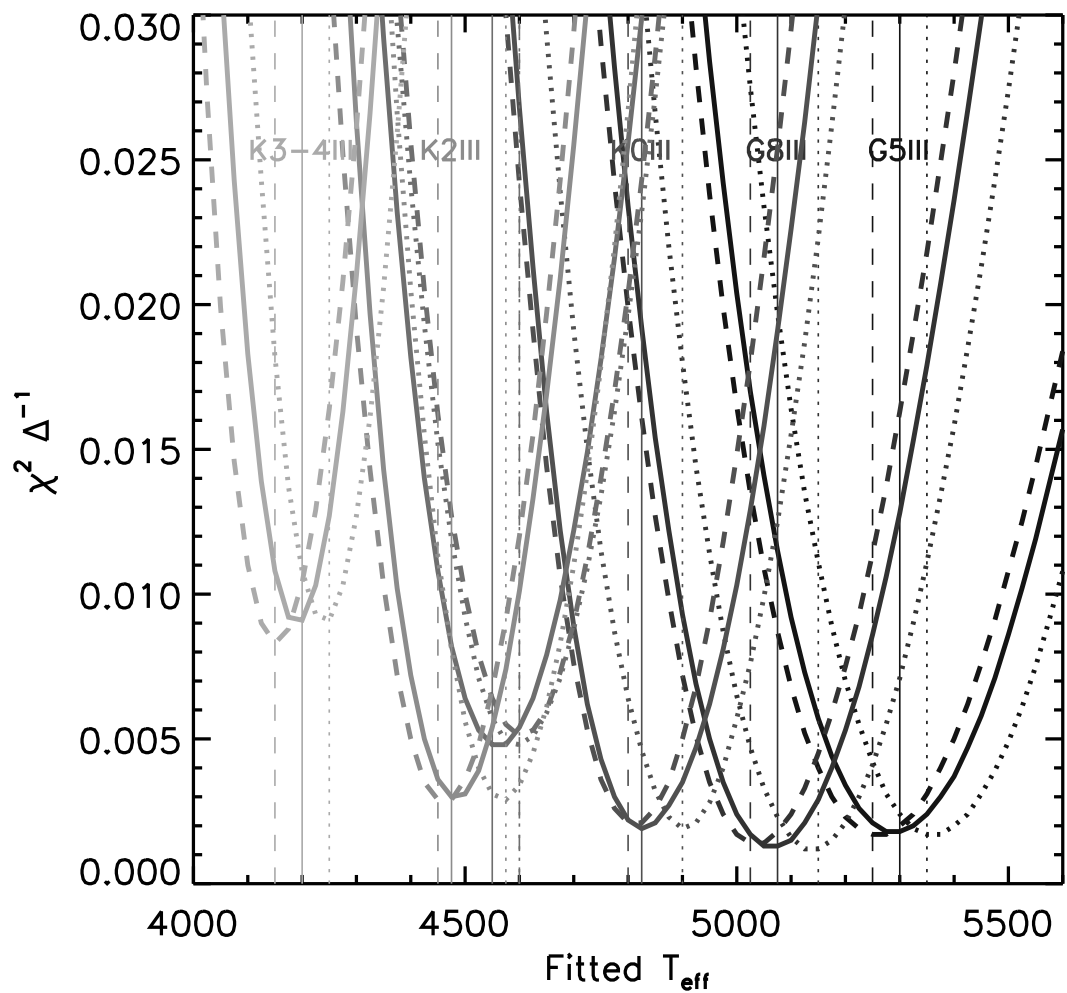}
\caption{The $\chi^{2} \Delta\lambda^{\rm -1}$ value as a function of trial $T_{\rm eff}$ value for
stars of $[{{\rm M}\over{\rm H}}]=0.0$.  We show $\chi^{2}$ for NLTE models with the GS98 and GASS10
abundances (solid and dashed lines, respectively).  The results for LTE models with the GS98
abundances are also shown for comparison (dotted line). 
  \label{refnltfitp00}}
\end{figure}

\begin{figure}
\plotone{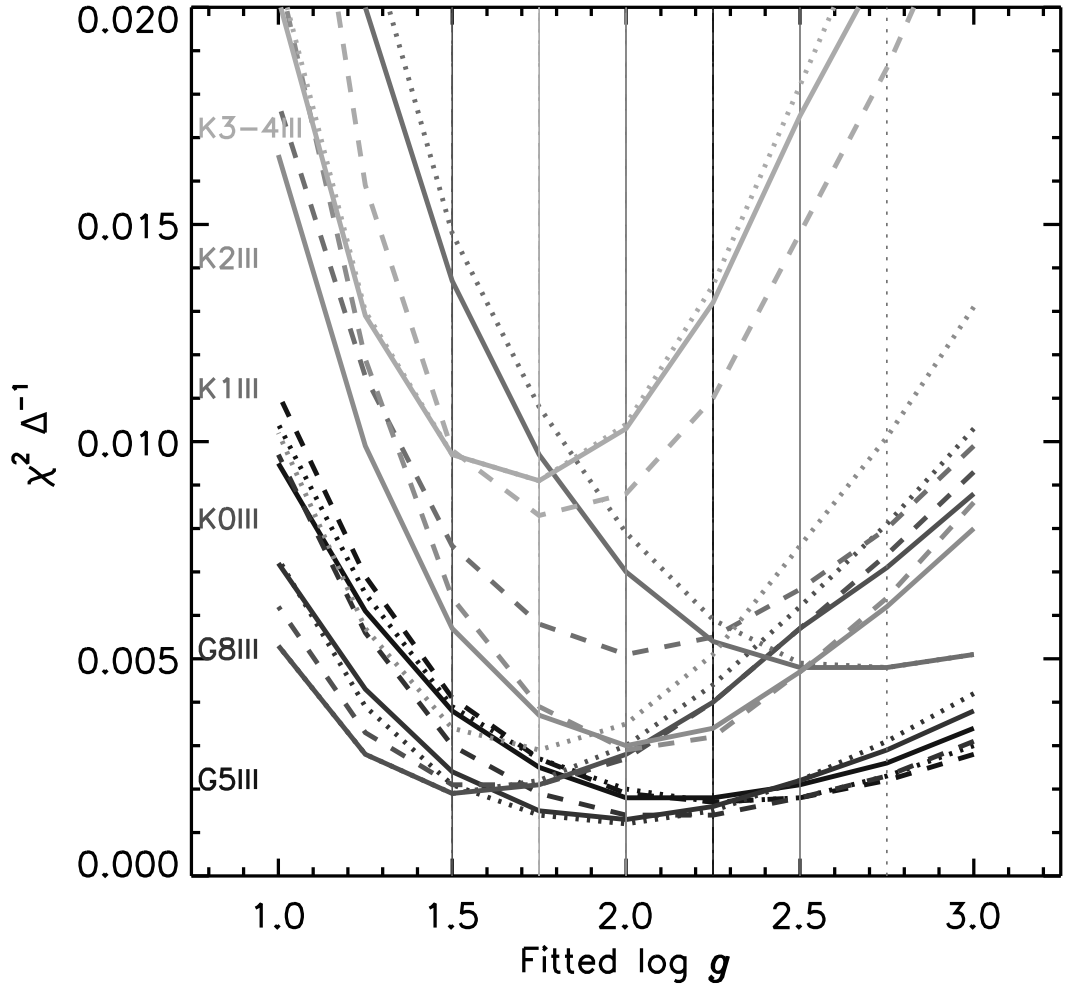}
\caption{Same as Fig. \ref{refnltfitp00}, but showing $\chi^{2} \Delta\lambda^{\rm -1}$
as a function of $\log g$.
  \label{refnltfitp00logg}}
\end{figure}

Table \ref{tabstatsbig} shows the best fit $T_{\rm eff}$ and $\log g$ values for the models of $[{{\rm M}\over{\rm H}}]=0.0$ 
for the GS98 and GASS10 abundance distributions in both LTE and NLTE.  
Figs. \ref{refnltfitp00} and \ref{refnltfitp00logg} show the variation of $\chi^2 \Delta\lambda^{\rm -1}$ with NLTE model $T_{\rm eff}$ 
and $\log g$, respectively, for the stars 
of solar $[{{\rm M}\over{\rm H}}]$ value for the models of GS98 and GASS10 abundance distributions, and Fig. \ref{reftempscalep00rm05} 
shows the best fit NLTE $T_{\rm eff}$ value as a function of $B-V$ color for these stars.  
Generally, the NLTE 
GASS10 abundances pervasively yield best fit $T_{\rm eff}$ values that are 25 to 50 K (one or two $\Delta T_{\rm eff}$ resolution elements) 
cooler than the older GS98 abundances, while yielding the same best fit $\log g$ values, a difference that we are marginally resolving. 
We conclude that there is evidence that adoption of the GASS10 abundances leads to best fit NLTE $T_{\rm eff}$ values for the blue band
and total SEDs that are 25 to 50 K cooler as compared to those with the GS98 abundances.  This is consistent with what we expect
from the comparison of GASS10 and GS98 SEDs discussed in Section \ref{sabund}.  Generally, the two abundance distributions
yield the same NLTE best fit $\log g$ value to within the $\Delta\log g$ grid resolution of 0.25.  
 
\paragraph{}
The one exception is the K1 III sample, for which
the GASS10 models yield a $T_{\rm eff}$ value that is 50 K {\it warmer}, while yielding a $\log g$ value that is 0.5 smaller (two
$\Delta\log g$ resolution elements). 
We note that the best fit $\log g$ value of the GS98 models (2.50) is higher than those
of the other spectral classes, for either abundance distribution, and that the value of $\chi^2$ for the $T_{\rm eff}/\log g$ pair 
4600/2.0 is identical to three 
digits after the decimal with that for the 4550/2.5 pair.  If the $\log g$ value is constrained to be 2.0 (GASS10 best fit model value) 
then the NLTE GS98 models yield a $T_{\rm eff}$ value of 4600 K, in closer agreement with the GS98 {\it vs} GASS10 behavior of the
other spectral classes. 

\paragraph{}

From Fig. \ref{reftempscalep00rm05}, which
also includes best fit LTE $T_{\rm eff}$ values, we note that the effect of NLTE on derived $T_{\rm eff}$ values compared to LTE is about the 
same for both the GASS10 and GS98 abundance distributions, namely a reduction of 50 to 75 K.  This is to be expected because
the GS98 and GASS10 distributions have identical values for the Fe abundance, and a brightening of the SED in the blue band with respect 
to the red band caused by NLTE over-ionization of \ion{Fe}{1} is the dominant NLTE effect in late-type stars (see \citet{rutten86}, \citet{shorth10}).  

\paragraph{$[{{\rm M}\over{\rm H}}]=-0.5$ }

\begin{figure}
\plotone{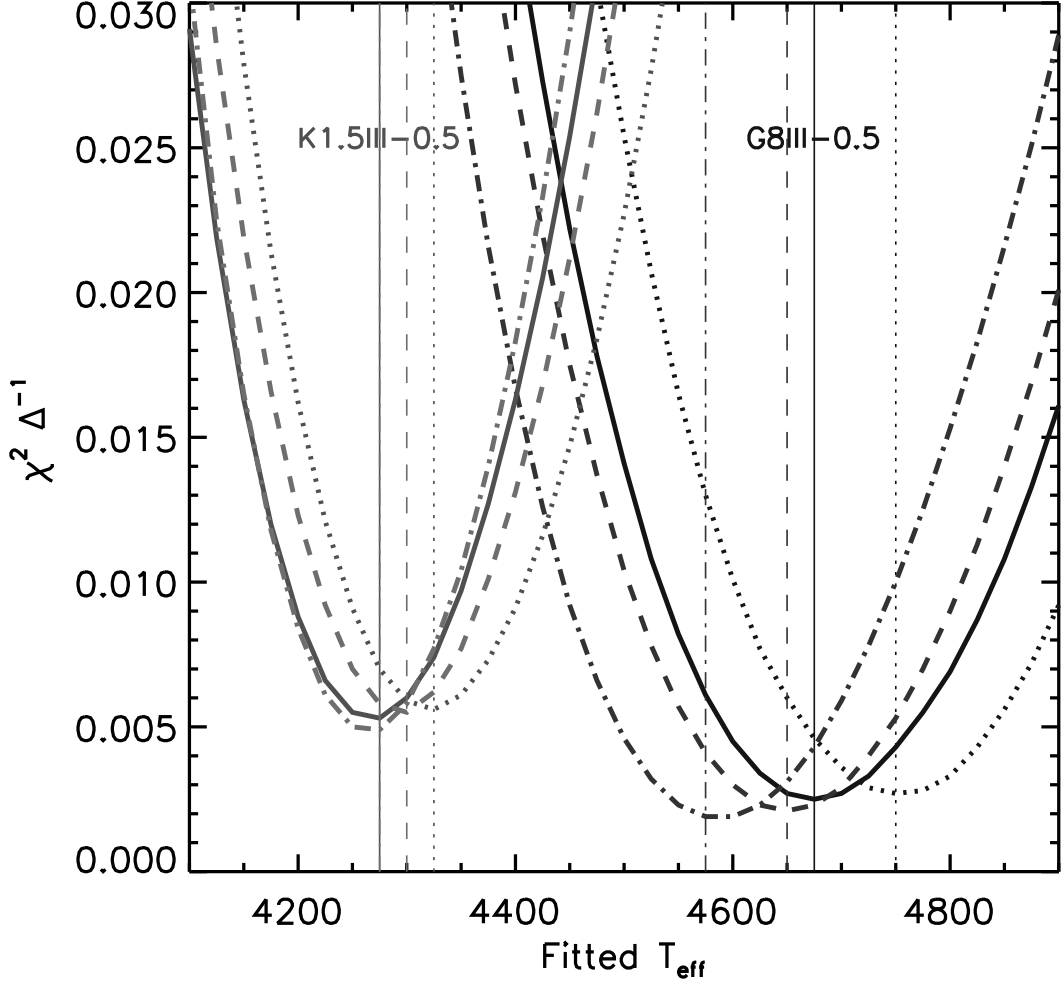}
\caption{Same as Fig. \ref{refnltfitp00} but for stars of $[{{\rm M}\over{\rm H}}]=-0.5$.  
We have included results for NLTE models with the $\alpha$-enhanced GASS10-$\alpha$
abundances (dot-dashed lines).  For clarity, we have omitted the NLTE models
with GS98-$\alpha$ abundances, however, the complete results in Table \ref{tabstatsbig} show
that the effect of $\alpha$-enhancement is similar for the GS98 and GASS10 models.  
As with Figs. \ref{refnltfitp00} and \ref{refnltfitp00logg},
we include the LTE models with GS98 abundances for comparison.
  \label{refnltfitm05}}
\end{figure}

 Table \ref{tabstatsbigm05} shows the best fit $T_{\rm eff}$ and $\log g$ values for the models of $[{{\rm M}\over{\rm H}}]=-0.5$ for all four relevant 
abundance distributions in both LTE and NLTE.  
Fig. \ref{refnltfitm05} shows the  variation of $\chi^2 \Delta\lambda^{\rm -1}$ with NLTE model $T_{\rm eff}$ for the stars of $[{{\rm M}\over{\rm H}}]$ equal 
to -0.5 for models of scaled GS98 and GASS10 distributions and the GS98-$\alpha$ and GASS10-$\alpha$ distributions, and Fig. \ref{reftempscalem05rm05}
shows the best fit $T_{\rm eff}$ values as a function of $B-V$ for both LTE and NLTE modeling.  For the G8 sample, adoption of the GASS10 abundances
decreases the derived $T_{\rm eff}$ value by 25 K, similarly to its effect on the $[{{\rm M}\over{\rm H}}]=0.0$ models, whereas for the K1.5 III sample, 
the GASS10 abundances {\it increase} the $T_{\rm eff}$ value by 25 K.  

\paragraph{$\alpha$ enhancement }

 More clearly, for the G8 sample, the effect of $\alpha$ enhancement is consistently to
further lower the inferred $T_{\rm eff}$ value, by 50 K for both the GS98 and GASS10 abundances.  By contrast, the effect of $\alpha$ enhancement 
on the K1.5 III sample is less clear; for the GS98 models it increases the derived $T_{\rm eff}$ value by 25 K, whereas its affect of the GASS10 models is to 
{\it lower} $T_{\rm eff}$ value by 25 K.  
Given that one numerical resolution element, $\Delta T_{\rm eff}$, is 25 K, we conclude that for stars of 
$[{{\rm M}\over {\rm H}}] = -0.5$, we are not resolving a 
clear trend for either the effect of choice of solar abundance distribution, or of $\alpha$ enhancement, with the possible exception of the effect
of the latter on the derived $T_{\rm eff}$ values of G8 III stars.  

\paragraph{}

The choice of solar abundance distribution has a negligible effect on the derived $\log g$ values.  However,
there is evidence that $\alpha$-enhanced models lead to derived $\log g$ values that are 0.25 to 0.5 (one to two $\Delta\log g$ resolution
elements) larger than those of scaled solar models, regardless of choice of solar abundance distribution.     

\paragraph{}
As with the models of $[{{\rm M}\over{\rm H}}]=0.0$, the NLTE reduction of the $T_{\rm eff}$ scale is independent of choice of solar
abundance distribution, although we note that the effect is more pronounced for the G8 III sample (75 K for three of the four abundance distributions),
than it is for the K1.5 III sample (25 K for three of the four abundance distributions, and never more than 50 K).

\paragraph{Blue band}

Fitting the ``blue'' band with LTE GS98 and GASS10 models results in best fit $T_{\rm eff}$ and $\log g$ values that 
are generally the
same as those from fitting the overall visible band, with occasional deviations by one numerical resolution element (25 K and 0.25,
respectively) higher or lower.  
The ``blue'' band and the overall visible band yield the
same stellar parameters to within our ability to resolve
differences.  This is not surprising because the heavily line
blanketed ``blue'' band is most sensitive to T${\rm _eff}$, and it thus
dominates the quality of the global fit.
For the NLTE models, GS98 and GASS10 models fit to the 
``blue'' band yield best fit $T_{\rm eff}$ values that are larger by 25 to 75 K (one to four resolution elements) than those fit
to the overall visible band for two of the six spectral class samples (G5 and K1). The larger ``blue'' band $T_{\rm eff}$ value tends to be 
accompanied by a best fit $\log g$ value that is lower by 0.25 to 0.5.  This result applies to samples of both metallicities. 
We conclude that stellar parameters inferred from fitting NLTE models to SEDs may be more dependent than LTE models on the wavelength 
region being fitted, and find that the effect depends on how heavily line blanketed the fitting region is, whether the fitting
region is to the blue of the Wien peak of the star's SED, or both.

\section{Comparison to other $T_{\rm eff}$ calibrations}

\begin{figure}
\plotone{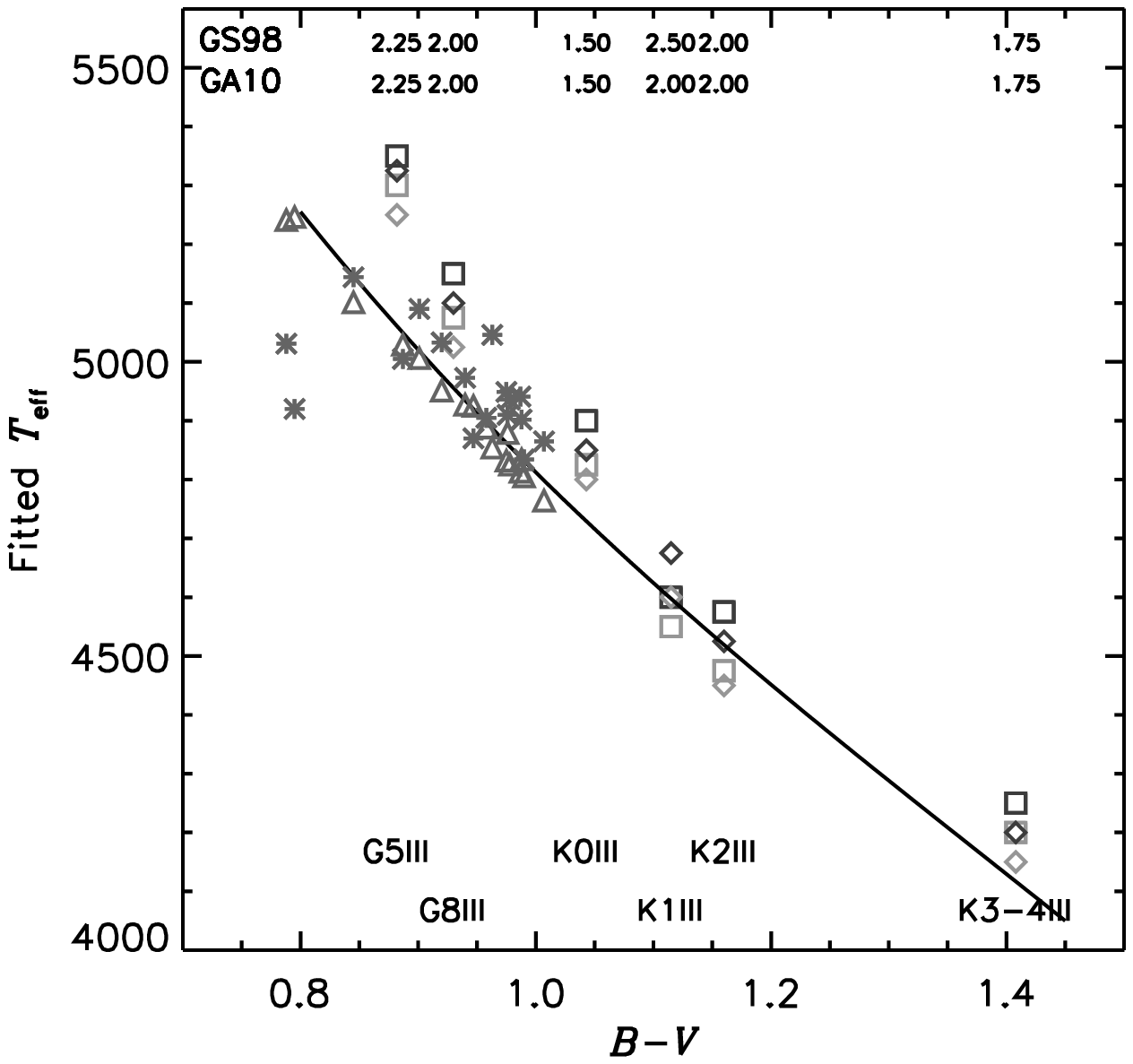}
\caption{Best fit T$_{\rm eff}$ values for stars of $[{{\rm M}\over{\rm H}}]=0.0$ from this investigation 
for NLTE (light gray) and LTE (dark gray)
models with the GS98 (squares) and GASS10 (diamonds) solar abundances.  For comparison we show the
results of other investigations of the $T_{\rm eff}(B-V)$ calibration: 
the IRFM calibration of \citet{ramirezm05} (solid black line) and values from \citet{wang11}
derived from $B-V$ photometry (triangles) and from \ion{Fe}{1}/\ion{Fe}{2} ionization equilibrium 
(asterisks).
Our spectral classes have been assigned corresponding $B-V$ values following the procedure described in 
Paper I to enable comparison with the other studies.  The annotation shows the best fit $\log g$ values 
derived from each solar abundance distribution (``GA10'' indicated GASS10 abundances). 
  \label{reftempscalep00rm05}}
\end{figure}

\begin{figure}
\plotone{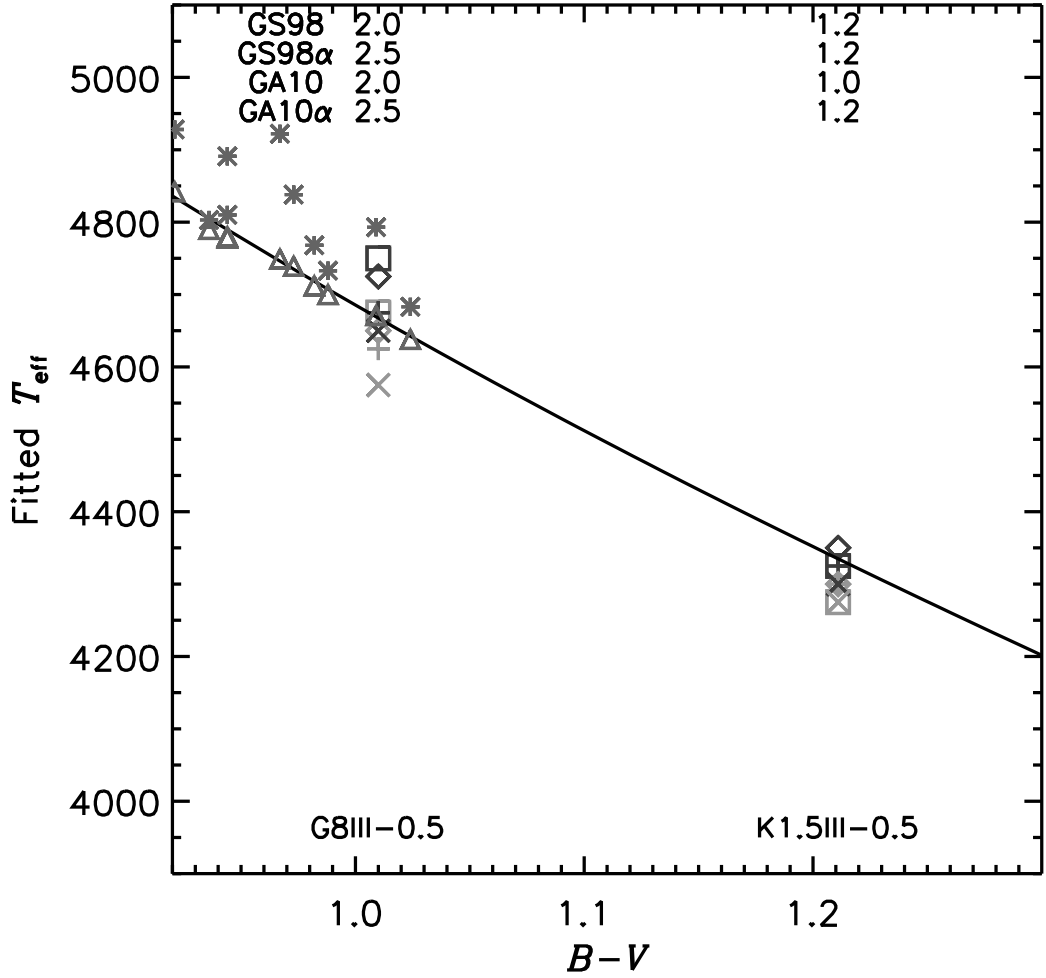}
\caption{Same as Fig. \ref{reftempscalep00rm05}, but for stars of $[{{\rm M}\over{\rm H}}]=-0.5$.
Results derived from the GS98-$\alpha$ and GASS10-$\alpha$ models are denoted with plus signs ('+') 
and 'X' symbols, respectively. 
The results of \citet{ramirezm05} and \citet{wang11} are those for the corresponding metallicity.
  \label{reftempscalem05rm05}}
\end{figure}

\begin{figure}
\plotone{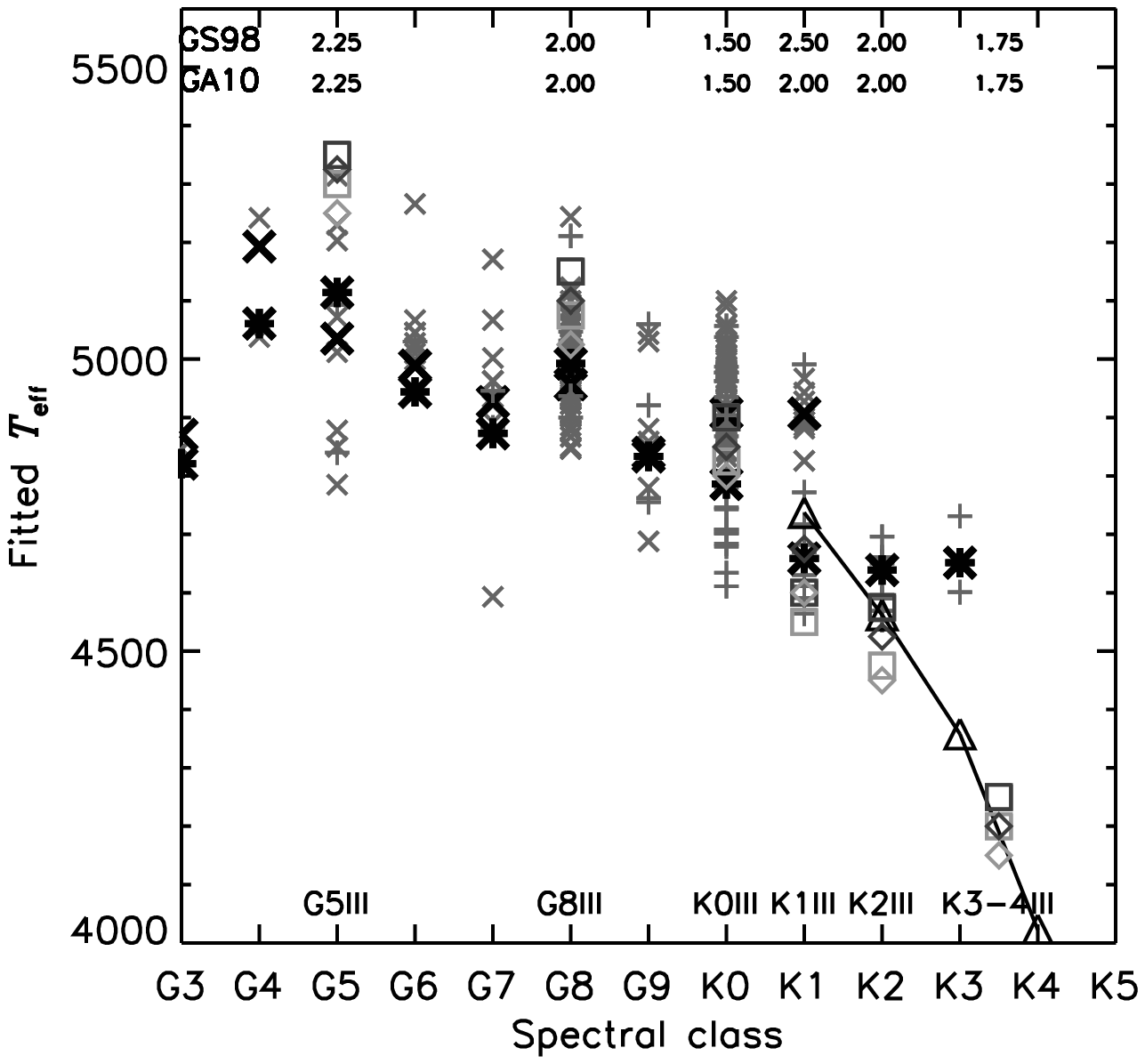}
\caption{Same as Fig. \ref{reftempscalep00rm05}, but with a comparison to the results of other
investigations of the $T_{\rm eff}$(spectral class) calibration: Interferometric results of 
\citet{baines10} (triangles), results of \citet{takeda08} from \ion{Fe}{1}/\ion{Fe}{2} ionization 
equilibrium (X's), and of \citet{mishenina06} from line depth ratios (asterisks).  For the
results of \citet{takeda08} and \citet{mishenina06} we also show the mean $T_{\rm eff}$ value
for each spectral class (large, dark X's and asterisks, respectively).
  \label{reftempscalep00b10}}
\end{figure}

Figs. \ref{reftempscalep00rm05} and \ref{reftempscalep00b10} show the comparison of our derived $T_{\rm eff}$ values with 
those of other investigators for the solar metallicity stars.  Papers I and II contains a review of the recent $T_{\rm eff}$ 
calibrations in the literature, based on a variety of complementary techniques, 
to which we compare our results.  To recapitulate briefly, \citet{ramirezm05} present a $T_{\rm eff}(B-V)$ 
calibration for a wide range of stars determined with the Infrared Flux Method (IRFM), \citet{wang11} present $T_{\rm eff}$ values for 
G giants determined from both
$B-V$ and \ion{Fe}{1}/II ionization equilibrium along with $B-V$ values, \citet{baines10} present $T_{\rm eff}$ values for 
select K giants determined from optical interferometry along with MK spectral classes, \citet{takeda08} present $T_{\rm eff}$ values
for G giants determined from modeling \ion{Fe}{1} and II lines for stars of given spectral class, and \citet{mishenina06} present 
$T_{\rm eff}$ values for G giants determined from line depth ratios for stars of given spectral class.  We note that Paper II contains
a thorough investigation of how well our NLTE $T_{\rm eff}$ values compare to the results of other studies with respect
to our LTE $T_{\rm eff}$ values.  Although we include the LTE results again here in Figs. \ref{reftempscalep00rm05} and \ref{reftempscalep00b10} 
for interest, we do not repeat the discussion of NLTE {\it versus} LTE. 

\paragraph{}

To facilitate the comparison to the results of \citet{ramirezm05} and \citet{wang11} in 
Fig. \ref{reftempscalep00rm05}, we determine
the mean observed $B-V$ color of our spectral class samples using the catalog of \citet{mermilliod91} following the procedure 
described in Paper I.  
In Fig. \ref{reftempscalep00rm05}, among the NLTE models, for every class except K2 III, NLTE models with the GASS10 abundance distribution 
yield $T_{\rm eff}$ values that are closest to, or are among the closest to, the values determined by \citet{ramirezm05} and \citet{wang11}.    

\paragraph{}

From Fig. \ref{reftempscalep00b10}, among the NLTE models, for two of our three spectral classes that overlap with the
interferometric results of \citet{baines10}, at the cool end of our spectral class range, the GS98 abundances 
lead to NLTE $T_{\rm eff}$ values that are in better agreement with that study.  For the G5 stars, at the other end of our 
range, the NLTE GASS10 abundances lead to $T_{\rm eff}$ values that are closer to the G giant results of \citet{takeda08}
and \citet{mishenina06}, although any conclusion is weakened because {\it all} our $T_{\rm eff}$ values are generally too high
compared to other results at the warm end of our $T_{\rm eff}$ range.

\paragraph{}

From Fig. \ref{reftempscalem05rm05} we can see that for the K1.5 sample, which consists entirely of multiple independent observations of Arcturus
($\alpha$ Boo, see Paper I), among the NLTE models, the GASS10 and GS98-$\alpha$ distributions give a $T_{\rm eff}$ value that is closest 
to the IRFM calibration of \citet{ramirezm05}.  That the GASS10-$\alpha$ distribution yields a more discrepant $T_{\rm eff}$ value
is note-worthy in that $\alpha$ Boo has been found to be $\alpha$-enhanced (see, for example, \citet{pdk93}).   

\section{Conclusions}

As expected, for solar metallicity stars, we find that the effect of NLTE on spectrophotometrically derived $T_{\rm eff}$ values 
compared to LTE is about the 
same for both the GASS10 and GS98 abundance distributions, namely a reduction of 50 to 75 K.
As with the models of $[{{\rm M}\over{\rm H}}]=0.0$, for the metal-poor stars the NLTE reduction of the $T_{\rm eff}$ scale is independent of choice of solar
abundance distribution, although we note that the effect is more pronounced for the G8 III sample (75 K for three of the four abundance distributions),
than it is for the K1.5 III sample (25 K for three of the four abundance distributions, and never more than 50 K).

\paragraph{}

For solar metallicity stars, we conclude that there is evidence that adoption of the GASS10 abundances lowers the inferred NLTE 
spectrophotometric $T_{\rm eff}$ scale by
25 to 50 K as compared to the GS98 abundances.  Generally, the two abundance distributions
yield the same NLTE best fit $\log g$ value to within the $\Delta\log g$ grid resolution of 0.25.  
For solar metallicity stars, NLTE models computed with the GASS10 abundances give, on balance, $T_{\rm eff}$ results that are 
marginally in better agreement with other T$_{\rm eff}$ calibrations in the literature determined by a variety of
means. 

\paragraph{}

We conclude that for stars of $[{{\rm M}\over {\rm H}}] = -0.5$, we are not resolving a 
clear trend for the effect of choice of solar abundance distribution on the $T_{\rm eff}$ value inferred from the SED. 
There is marginal evidence, based entirely on the G8 III sample, that the adoption of $\alpha$-enhancement leads to
a reduction of 50 K in the NLTE SED-fitted $T_{\rm eff}$ value.  
The choice of solar abundance distribution has a negligible effect on the derived $\log g$ values.  However,
there is evidence that $\alpha$-enhanced models lead to derived $\log g$ values that are 0.25 to 0.5 (one to two $\Delta\log g$ resolution
elements) larger than those of scaled solar models, regardless of choice of solar abundance distribution.  With 
only two spectral classes, it is difficult to draw conclusions about which abundance distribution better
fits other $T_{\rm eff}$ calibrations.   

\paragraph{}

We conclude that stellar parameters inferred from fitting NLTE models to SEDs are more dependent than LTE models on the wavelength 
region being fitted, and find that the effect depends on how heavily line blanketed the fitting region is, whether the fitting
region is to the blue of the Wien peak of the star's SED, or both. 

\paragraph{}

We will make both the library of observed SEDS and the NLTE (and corresponding LTE) grid of model
SEDS with all abundance distributions available to the community by ftp \\
(http://www.ap.smu.ca/ ~ishort/PHOENIX).



\acknowledgments

CIS is grateful for NSERC Discovery Program grant 264515-07.  The calculations were
performed with the facilities of the Atlantic Computational Excellence Network (ACEnet).

\clearpage



\clearpage

\clearpage

\clearpage

\clearpage

\clearpage

\clearpage

\clearpage

\clearpage

\clearpage

\clearpage

\clearpage

\clearpage

\begin{deluxetable}{lrrrrrrrr}
\tablecolumns{9}
\tablecaption{Best fit parameters ($T_{\rm eff}$/$\log g$) for total and ``blue''-band SEDs: Models of $[{{\rm M}\over{\rm H}}]=0.0$ }
\tablehead{
\colhead{} & \multicolumn{4}{c}{LTE}                               & \multicolumn{4}{c}{NLTE}                     \\ 
\colhead{} & \multicolumn{2}{c}{GS98} & \multicolumn{2}{c}{GASS10} & \multicolumn{2}{c}{GS98} & \multicolumn{2}{c}{GASS10}                    \\ 
\colhead{SED:} & Total & Blue & Total & Blue & Total & Blue & Total & Blue \\ 
} 
\startdata
G5 & 5350/2.25 & 5375/2.25 & 5325/2.25 & 5325/2.25 & 5300/2.25 & 5325/2.00 & 5250/2.25 & 5275/2.25 \\
G8 & 5150/2.00 & 5150/2.00 & 5100/2.25 & 5075/2.25 & 5075/2.00 & 5075/2.00 & 5025/2.00 & 5025/2.00 \\
K0 & 4900/1.50 & 4900/1.50 & 4850/1.75 & 4850/1.75 & 4825/1.50 & 4825/1.50 & 4800/1.50 & 4800/1.50 \\
K1 & 4625/2.50 & 4575/3.00 & 4675/2.00 & 4700/1.75 & 4550/2.50 & 4625/2.00 & 4600/2.00 & 4675/1.50 \\ 
K2 & 4575/1.75 & 4575/1.75 & 4525/2.00 & 4525/2.00 & 4475/2.00 & 4475/2.00 & 4450/2.00 & 4450/2.00 \\
K3-4&4250/1.75 & 4250/1.75 & 4175/2.00 & 4175/2.00 & 4200/1.75 & 4200/1.75 & 4150/1.75 & 4125/2.00 \\
\hline\\
\enddata
\label{tabstatsbig}
\end{deluxetable}

\begin{deluxetable}{lrrrrrrrr}
\tablecolumns{9}
\tablecaption{Same as Table \ref{tabstatsbig}, but for total SED fits only: Models of $[{{\rm M}\over{\rm H}}]=-0.05$ }
\tablehead{
\colhead{} & \multicolumn{4}{c}{LTE}                               & \multicolumn{4}{c}{NLTE}                     \\ 
\colhead{} & \multicolumn{2}{c}{GS98} & \multicolumn{2}{c}{GASS10} & \multicolumn{2}{c}{GS98} & \multicolumn{2}{c}{GASS10}                    \\ 
\colhead{} & Solar & $\alpha+$ & Solar & $\alpha+$ & Solar & $\alpha+$ & Solar & $\alpha+$ \\ 
} 
\startdata
G8 & 4750/2.00 & 4675/2.50 & 4725/2.00 & 4650/2.50 & 4675/2.00 & 4625/2.50 & 4650/2.00 & 4600/2.25  \\
K1.5&4275/1.50 & 4325/1.25 & 4350/1.00 & 4300/1.25 & 4275/1.25 & 4300/1.25 & 4300/1.00 & 4275/1.25  \\
\hline\\
\enddata
\label{tabstatsbigm05}
\end{deluxetable}

\end{document}